\input amstex
 \loadmsam \loadmsbm
\input amssym.tex
\def\[{\lbrack}\def\]{\rbrack}
\catcode`\[=1\catcode`\]=2\catcode`\{=12\catcode`\}=12 

\magnification=\magstep1\baselineskip=24pt
\loadeusm

\def\foo#1#2[\baselineskip=10pt\plainfootnote[$^[#1]$][#2]\baselineskip=24pt\thinspace]
\def\foop#1#2[\baselineskip=10pt\plainfootnote[$^[#1]$][#2]\baselineskip=24pt\ ]
\def\foow#1#2[\baselineskip=10pt\plainfootnote[$^[#1]$][#2]\baselineskip=24pt]
\def\instert[[\it Creating Modern Probability], Cambridge, 1994]

   \centerline[The Logical Status of Phyical Probability Assertions]
\centerline[by Joseph F.\ Johnson]

\
In this paper we offer a definitive solution to the major problem in
that part of the philosophy of probability which is primarily logical, the
question of the meaning of probability.  We abstract severely from all
epistemological considerations, we do not adopt a positivistic approach,
and the applications to the problems of quantum mechanics are dealt with 
separately in another paper\foo 1  [`Thermodynamic Limits, Non-commutative 
Probability, and Quantum Entanglement' in, 
Quantum Theory and Symmetries III, Cincinnati 2003, ed. by Argyres et al, Singapore, 2004, pp.133-143.] 
using the earlier approach of von Plato\foo 2
 [``Ergodic Theory,'' in Skyrms, Harper, eds., [\it
Causation, Chance, and Credence], [\it Proceedings of the Irvine Conference on
Probability and Causation, 1985] vol.\ 1, Dordrecht, 1988 pp.\ 257-277.] 
 instead 
of the present approach.  This approach as well  
as von Plato's seems to allow one to solve the problem of scientific induction 
as well, this is reserved for an intended sequel.  
It is well known that each
of the three main rival theories of the meaning of probability, the frequency
theory, the Keynesian theory, and the de Finetti-Savage (`making book')
theory, have fatal drawbacks\foo 3
[For example, the frequency theory has been acutely criticised by
Kolmogoroff in his contributed chapter to Aleksandrov, Kolmogorov, and Lavrent'ev,
ed.s, [\it Mathematics its content, Methods, and Meaning], 2nd ed., Moscow,
1956, we cite from a Cold War translation published in 1963 in Cambridge,
Mass., p.\ 239, ``\dots it is clear that this procedure will never allow us to be
free of the necessity, at the last stage, of referring to probabilities in the
primitive imprecise sense of this term.''] and so there is literally no consensus
among philosophers as to the status of this problem.  (That decades later
Kolmogoroff went off on a tangent and evolved his own idiosyncratic theory
of the logical status of probability statements is beside the point.)    
               
     In particular the attempt to analyze the logical
structure of either propositions asserting probability or
propositions ``asserted" probabilistically, has not met with the
same success as, say, Frege and Russell's (or Wittgenstein's)
analysis of the propositions of mathematical logic (or physical
science).  The aim of this paper is to show that this is because
the analysis is not in the least parallel.  We will propose a
logical analysis of probability assertions which takes its cue from
recent developments in statistical mechanics and deterministic
chaos.  Therefore, assertions of the probability of a state of
affairs are to be unpacked as having reference to an (implicit)
approximative model of reality---in contrast to normal
propositions, which refer to reality directly.  This view is almost
forced on one if one recasts the mathematical physics of today on
statistical mechanics into the framework of Russell-Whitehead and
Bourbaki.

     We will accomplish this by taking into account Kolmogoroff's critique of the
logical circularity of the frequency theory, otherwise the most attractive
candidate.  We can remove this circularity by passing to the limit, but of
course we recognise that `limit' is not a word with precise meaning in general,
so we have to specify what kind of limit.  Kolmogoroff's critique\foo 1 [[\it
op.\ cit.,] p.\ 253f., he remarks about vague talk of limits, ``Such definitions
roughly correspond to the `definition' in geometry of a point as the result of
trimming down a physical body an infinite number of times, each time decreasing
its diameter by a factor of 2.''] can be met by the Darwin-Fowler concept of the
thermodynamic limit.

     The philosophy of probability has many aspects, 
but we choose to focus on the philosophy of science and
probability within the realm of its usage in physics. 
This paper will concentrate on the purely logical
problem of the meaning of a statement that such and such physical event's
occurrence has, had, or will have, probability x where x is a real number in
between 0 and 1.  We will abstract from all epistemological questions and all
questions of what is the use of such a statement, except that a few remarks will
immediately follow from our solution of the purely logical problem of meaning. 
This seems to solve the problem of induction (following Wiener's  
ideas about induction and his theory of prediction). \foo 1 [Wiener, [\it
Extrapolation, Interpolation, and Smoothing of Stationary Time Series With
Engineering Applications], p.\ 47: ``This allows us to
 identify averages made on the observable past of a time series with averages
to be subsequently obtained from the now unattainable future\dots It is this
step which constitutes the logical process of induction.'']

The point of this paper is to completely subsume the meaning of such a
statement within the Russell-Wittgenstein, Bourbaki framework of mathematical
logic as applied to physical science.  This is reductionist, but it works.  All
the usual scientific uses of the concept of probability and all the
mathematical theorems about it are validated by this philosophical
interpretation, obviously a desideratum of any such, and no new philosophical
concepts or difficulties are raised: the fatal criticisms of
frequentism are remedied.

Now I wish to emphasize that the point of this paper is to
successfully complete an experiment.  Can a purely deterministic,
classically logical, reduction of the notion of probability be made?  This
requires a rather severe focus.  Only immanent objections are relevant to
this question, which is not exactly the same as the question of whether this
account of the meaning of probability is `true.'  Of course it has to
account for the uses made of it in science and apply to individual events, or
else it is not the `notion of probability.'  This said, success along these
lines renders the rival theories rather pointless: historically, their
motivation was always the perceived failures of such experiments as this one.

Of the three main rivals, and other less important theories, as to the
logical meaning of probability statements,the frequency theory is the most
attractive but the most logically flawed.  Nevertheless, it will be our starting
point since most scientists go ahead and use it without worrying about its
problems, and this paper will, in a way, justify that.  The frequency theory is,
of course, incapable of being stated coherently without involving a logical
vicious circle, as Kolmogoroff famously pointed out.  I would like to call this
the Rosencrantz and Guildenstern effect, after the opening scene of Tom
Stoppard's play.

Rosencrantz, or perhaps it was Guildenstern, opens the play by tossing a fair
coin sixty-four, or perhaps it was sixty-five, times in a row and getting the
same result---heads, or perhaps it was tails---every time (but the last).  The
probability that any given toss gives a given result is one-half.  The
probability that the play-opening tosses-event would occur is small but
positive.  Evidently the meaning of the statement `the probability that this
coin yields heads is fifty per cent.' does not imply that the frequency with
which heads occurs in any finite set of trials will be anything in particular. 
This is a problem with the frequency theory.  It is circular to say that
nevertheless, it asserts that the probability that the expected frequency will be
attained or departed from is what the calculus of probabilities allows us to
calculate it to be, because this involves an endless regress.  It is simply
false, due to mathematical ignorance, to assert that in an infinite number of
trials the frequency will definitely be fifty per cent., because an infinite
number of trials gives us an infinite probability space, in fact a continuum of
possibilities, and so events with zero probability are not impossible.  It is
entirely possible that every toss of an infinite sequence of tosses of
a fair coin (conceding the physical meaningfulness of this for argument's sake)
will be heads.  To bring up its probability is again a circular regress.

Of course there is some validity to the intuition that, `in the limit' the
expected frequency will be obtained and that this is what we mean when we say
that the probability that a toss will be heads is one-half.  But as we saw
above, it does not mean that it is impossible that the expected frequency will
fail to obtain, and further, no one has satisfactorily defined what sort of
limit.  The word limit has no physical or mathematical meaning in isolation. 
There are various kinds of limits, one for each context, and they have various
properties.  This paper will remedy the problem identified by Kolmogoroff, the
Rosencrantz and Guildenstern effect, and will give a logically well-defined
meaning to the intuition `in the limit' by introducing the logically rigourous
treatment of the thermodynamic limit pioneered by Darwin and Fowler,\foo 1
[Fowler, [\it Statistical Mechanics], Cambridge, 1929.] and applied to Brownian
motion by Wiener and Kolmogoroff, as exposited in the book by Khintchine, [\it
The Mathematical Foundations of Statistical Mechanics], New York, 1949.  There
are some similarities to a proposed theory of the meaning of probability by
Professor Jan von Plato\foo 2 [``Ergodic Theory,'' in Skyrms, Harper, eds., [\it
Causation, Chance, and Credence], [\it Proceedings of the Irvine Conference on
Probability and Causation, 1985] vol.\ 1, Dordrecht, 1988 pp.\ 257-277.] of
Helsinki University, and there are some similarities to one of the many
contradictory proposals thrown out by Aristotle in his discussions of chance and
happenstance.  Probability statements require to be unpacked in a way analogous
to Russell's famous theory of descriptions, but vastly more complicatedly, and we
will argue that their logical structure is quite deeper than their superficial
grammar.  This means that `probability,' like `variable' and `the' is a
linguistic artifact or facon de parler, and the statements in which it occurs can
always be unpacked into unstraightforward but reducible determinate 
statements about the physical world.

It is well known from classical statistical mechanics, especially the theory of
Brownian motion\foo 1[Heims, Steve, [\it John von Neumann and Norbert Wiener:
From Mathematics to the Technologies of Life and Death], Cambridge, Mass., 1980,
pp.\ 72-78 is a nice popularisation.] that stochastic models are useful
approximations to deterministic models: indeed, the one can approximate the other
to within any desired degree of approximation, so that experimentally they are
indistinguishable.  The logic of approximate statements, i.e., propositions that
such and such a quantity is definitely inside a precise and determinate range of
possibilities, is unglamourous precisely because nothing new except measure
theory is introduced into philosophy or science, but it proves sufficient to
answer all the puzzles about probability.  Indeed, such approximate statements
are simply precise propositions on the same logical footing as the usual
propositions about points and exact equalities, and this has always been felt. 
(The dictionary definition of `centum' includes such an approximate meaning quite
explicitly: when, too, we talk in round numbers, this convention is understood. 
There is nothing new or revolutionary here and no changes in classical logic are
required.)  Then, too, the fact that the calculus of probabilities has already
been successfully incorporated within the classical mathematical logic without
any changes needed is a sort of omen that the philosophy might follow the same
course.  

One of Aristotle's many suggestions about happenstance is that it arises when
one deterministic system interacts with another deterministic system and there
exists no over-arching combined system which keeps track of the interaction
terms and is deterministic.  Evidently he was not familiar with Jacobi's work
on dynamics.  Ever since Jacobi's time, a revolutionary change in our attitude
to combined systems had occurrred.  It used to be taken for granted that one
could not unify all systems into an overarching dynamic.  But Jacobi's work
showed that in mechanics, at least, one could practically formalise the
process.  Slowly the belief in a grand unified theory has taken over the
physical sciences because of the compelling vision that there will always be an
interaction term which takes it all into account.

Nevertheless, there is one barrier which must remain:  the interaction between
an individual particle, such as the dust mote in Brownian motion, and a
Hamiltonian heatbath, must be resistant to this Jacobian imperative: the
heatbath is not a dynamical system at all, but is a thermodynamic limit.  The
barrier between the finite and the infinite clearly marks off the realm of
dynamics from another realm.  For two reasons:  firstly, a finite system is a
dynamical system in the technical sense of the word: there is a symplectic phase
space as is usual in Hamiltonian mechanics, Liouville measure, invariant under
the dynamical flow of states, etc.  But even though physicists have various ad
hoc ways of mathematically modelling infinite systems, they are not dynamical
systems in the same sense.  Secondly and more important from a logical point of
view (since after all one could wonder whether the first, technical barrier
could not be overcome by suitable generalisation of the definition of dynamical
system), `the' thermodynamic limit is not an object.  The logically rigourous
treatment of the process of `passing to the thermodynamic limit' (below) shows
that it is not analogous to, say, passing to the limit of a sequence of polygons
and obtaining a circle.  The linguistic similarity is highly misleading,
although it does form the working basis for the typical sloppy undergraduate way
of thinking of statistical physics.  In this sense, then, this paper offers a
validation of Aristotle's insight.  Probability is a linguistic phenomenon that
arises when we wish to discuss the interaction, as in Brownian motion, between
two systems which cannot be subsumed under a common overarching framework of the
same dynamical type as each of them separately.

     The ignorance interpretation of probability has never been
satisfactorily interpreted in the (above mentioned) framework of
logic.  Naturally enough, for it is observer dependent and not
linguistically based.  It is not a feature of language.  The
interpretation we will put forward is linguistic, although
contextual:  It depends not only on the explicit proposition, but
a model (which is a set of propositions).  The ignorance theory
also does not allow for real randomness, and so is rejected by
Quantum Mechanics.  Our theory will allow the expression of real
randomness---as an approximation to a deterministic reality.  No
logical theory can very satisfactorily allow real randomness---as
a very matter of reference, because of the following objection.
     
Suppose the logic allows assertions $q$ such as `$p$ has probability
$\frac12$,' $p$ being a physical event.  What of the meta-language?  The
suggestion that there are meta-propositions about $q$ which are
random is evidently absurd.  But the resulting inhomogeneity
between a probabilistic logic governed by a fully classical
meta-logic is rather unsatisfactory.  The natural conclusion would
be that the stochasticness of the logic is generated by the
physicality of (some of) its objects.  But this would leave the
logic without any role in coming to grips with the stochasticness.
     
     For this reason, indeed, Kolmogoroff has been able to show
that the mathematical structure of probability theory is completely
ordinary and can be reduced to (classical) mathematical logic (say,
\it Principia Mathematica\rm).  Of course it is understood by all that the
mathematical definition of probability is merely a synonym, and not a
philosophical definition of probability.

     Since our proposed interpretation of probability propositions
can account for the apparent stochasticness of reality, it would
seem, then, that there is no need for stochasticness in the logic
to be generated by physics.  This justifies the superficial
observation that a probability proposition must still, after all,
be a certain assertion of its meaning.  Before proceeding, it is important to
stress that the probabilities of classical statistical mechanics do not arise
out of ignorance, hidden variables, or coarse-graining and that these concepts
are not mentioned at all in this paper.  Hence, the meaning of probability has
nothing to do with these sloppy undergraduate ways of approaching
thermodynamics.

     Our thesis is that the concept of probability is an artifact
of a certain approximation procedure.  It is thus neither
linguistic (or logical) nor physical. It is only mathematical.  We
must begin by discussing the logical structure of statistical
mechanics in a way that does not involve the notion of probability. 
This structure is now apparent; the belief that it can be proved
vigorously in the imaginable future is current, \it pace \rm von Mises,
even though this is not accomplished yet.

     We discuss the concrete case of classical Brownian motion, for
which the rigorous results are in place.  When Einstein gave such
an impetus to the theory of Brownian motion, it was conceived
classically, so we flesh out a Newtonian picture for our
discussion.  Suppose that in a closed region (with reflective
walls) we have  $n$  small billiard balls and one large sphere, much
larger than the balls.  Call this dynamical system  $X_n$  and its
phase space  $V_n$  of all possible positions and velocities of all
the balls and the sphere.  Then  $V_n$ is  $6(n + 1)$-dimensional. 
For a fixed total energy of  $X_n$  only a hypersurface $E_n\subset V_n$ is
relevant.

     The problem is to describe the motion of the sphere,  $S_n$.  To
make this well defined, one must specify the initial conditions of 
$X_n$  at time  $t=0$, call this point in  $E_n$,   $v_0$.  Then let 
$X(t,v_0)$ 
 be the resulting path of  $S_n$  in the system  $X_n$ with initial
conditions  $v_0$, regarded as a function of $t$ for fixed  $v_0$.  It is
impractical to solve this problem exactly.  Nevertheless, this is
a deterministic problem, so for each  $v_0$, $X(t, v_0)$  is a function
of   $t$.  The set ${X(t,v_0)|v_0\in E_n}$ is called the space of ``sample
paths.''  It is a set of functions.  Since each function has the physical
interpretation of a ``path,'' it can be interpreted as a set of paths.

     The procedure of Einstein and Wiener was rather to obtain an
approximate solution to this problem, valid if  $n$  is very large. 
Wiener constructed a different space,  $W =
{X(t,\alpha)}_[-\infty<t<\infty,\ 0\leq\alpha\leq1]$   the
Wiener process.  For 
fixed   $\alpha$, $X(t,\alpha)$ is a
continuous function  (such that $X(0)=0$,
 as can be assumed).  Every such continuous function arises as an 
$X(t,\alpha)$   for some   $\alpha$.  We put Euclidean (Lebesgue) measure on the
interval  $0\leq\alpha\leq1$  and transfer this to $W$, so $W$ becomes a
probability space in the sense of Kolmogoroff.  (Wiener showed that as such,
the   $X(t,\alpha)$  satisfied the properties of Einstein's
model---Einstein had treated it postulationally, not constructively.)

     The intuitive way people think about this is (illogical---we
will soon replace it with a rather different way) as follows. 
The  $\alpha$  parameter is thought of a analogous to  $v_0$.  The $W$ is the
limit of  $V_n$ as  $n$  approaches infinity (`in the thermodynamic
limit'---i.e., the energy and pressure are held constant so the
mass of each billiard ball shrinks proportionately).
For each  $v_0$,  $X(t,v_0)$ is differentiable almost everywhere in
$t$. (I.e., everywhere except at collisions with the balls.)  But as
we imagine the billiard balls to get smaller but more numerous, the
collisions are more frequent.  So, corners or kinks (of
non-differentiability) in the path become more frequent.  In the
limit, they are almost everywhere.  Thus, almost all the $X(t,\alpha)$  are
(continuous but) non-differentiable almost everywhere in $t$.

     Now in the limit-model, the stochastic model, physicists tend
to think intuitively of each sample path (i.e., each  $X(t,\alpha)$       for
fixed  $\alpha$) as the motion of a sphere under ``random impacts" from
the surrounding medium---this was Einstein's view.

     The thesis of this paper is rather that the probabilistic
aspects are purely artifacts of $W$.  Now $W$ only has validity as a
mathematical approximation device for studying certain aspects of
$V_n$ when  $n$  if very large.\foo 1 [Cf.\ Wiener, ``Logique, Probabilite
et Methode Des Sciences Physiques,'' Toutes les lois de probabilite connues
sont de caractere asymptotique\dots les considerations asymptotiques n'ont
d'autre but dans la Science que de permettre de connaitre les proprietes des
ensembles tres nombreux en evitant de voir ces proprietes s'evanouir dans la
confusion resultant de las specificite de leur infinitude.  L'infini permet
ainsi de considere des nombres tres grands sans avoir a tenir compte du fait
que ce sont des entites distinctes.''  That is, it is a method, not an object.] 
It is only  $V_n$  that is a positivistic model of physical reality in the sense
that there is a correspondence between objects of  $V_n$  and physical objects,
between relations (or structures, such as functional dependence) in $V_n$ (between
objects of $V_n$) and physical relations between physical objects.  The model 
$W$ is only about reality indirectly:  it is `about'  $V_n$  which is about
reality.  (Strictly speaking, it is not even about  $V_n$:  arguably, it is not
really a model, since it is not about anything.)

     Some questions that can be asked about  $V_n$  can be
approximately answered by studying  $W$---the physical question of
Einstein's original series of papers, for example.  Other features
of  $V_n$  are not preserved by the limiting process and so are not
present in  $W$.  For example,  $V_n$  satisfies Poincare recurrence for
each  $n$.  But  $W$  does not.  A rigorous treatment of this
approximation procedure is still lacking in
general statistical mechanics.  It has been done by Ford-Kac-Mazur\foo 1 [Ford,
Kac, Mazur: ``Statistical Mechanics of Assemblies of Coupled Oscillators, [\it
Journ.\ Math.\ Phys.\ ] vol.\ 6 (1965), pp.\ 504-515.] (although not exactly for
billiard balls and a sphere, but rather for phonons and a harmonic oscillator),
although a complete characterisation of all those properties of  $V_n$  which
smoothly approach a limit property of  $W$ is still lacking.  But enough has been
done to render unlikely von Mises's flat statement that a justification of
statistical mechanics is inconceivable.

The following is in its logical, modellic structure, essentially due to Darwin
and Fowler.  The details of their proofs were simplified by Kolmogoroff and,
independently, Wiener, and exposited by Khinchine ([\it loc.\ cit.])\rm 

     Let  $P$  be a propositional function of models such that for
all  $n$,  $V_n$  is in its domain.  But  $W$  need not be.  By saying
that  $P$  can be approximately answered by a study of  $W$  we mean
that  $P$  is of the form $|f|=0$       where  $f$ is a function of  $V_n$  with
values in a metric space (or, for simplicity, we may assume  $f$  is
a numerical function) and the proposition  $Q$, ``for every  $\varepsilon>0$   
there exists  $N$  such that for all $n>N$ we have 
$|f(V_n)|<\varepsilon$''         is equivalent to a proposition  $Q'$  about 
$W$.  (The words ``study" and ``about" are vague, and correlatively so.  The
original statement, ``$P$ can be approximately answered by a study of  $W$" is
not a formal proposition.  If, however,  $V_n$  is some sort of function of an
additional parameter, say  $H$, and so is  $W$, then we can ask, for
example, that for  $P$ independent of  $H$,  $Q'$  should be also. 
Similarly if  $P$  depends on $H$ only through algebraic operations on 
$f$, etc.  These are examples of how to make ``study" and ``about" have
a precise meaning.)

     For example, we could take  $P$  to be ``The average motion of
the sphere after ten seconds (at a fixed energy, etc.) where the
average is taken with respect to a (given) weighting of initial
conditions  $v_0\in E$, is $\sqrt[10]$."  Then  $Q'$ is
``$\int_0^1X(10,\alpha)\,w(\alpha)\,d\alpha$      is 
    $\sqrt[10]$" where  $w$  depends on the weighting mentioned in  $P$.  We 
can also let  $E$ and the relative sizes of the balls and sphere, etc.,
vary and obtain families of  $P$  and  $Q'$.

     Note that the intuitive view of  $X(t,\alpha)$       cannot be formalized: 
Since the Wiener process is not in fact a dynamical system. 
Admittedly, $X(t,\alpha)$  satisfies a so-called stochastic differential
equation.  But an analysis of the linguistic structure of this
reveals its utter lack of parallelism to  $V_n$, or any  $X_n(t,v_0)$. 
This follows:  for fixed  $t$, $X_t(\alpha)=X(t,\alpha)$ is obviously a function
of  $\alpha$  (with values in a three dimensional space of vectors).  And 
$\alpha$   has as its domain a probability space.  In fact, $X_t$     is what is
called a random variable.  The definition of a stochastic
differential equation is far removed from that of a differential
equation, but even granting part of the intuitive view (which we
need not), which follows, we still will carry our point.

     An ordinary differential equation is thought of as a way of
relating a variable at one point in time to a variable at a
neighboring point in time (by means of a limiting process)---similarly for
stochastic differential equations, in particular the one which the Wiener
process  ${X_t}_[-\infty<t<\infty]$              satisfies.  (But this parallel
is specious only to those who are ignoring the fact that a so-called ``random
variable" is not a variable.  A random variable is an object, but a variable is
a linguistic construction (hence it could be an object only for the
meta-language).  In particular, a random variable is a measurable function from
a probability space to the real numbers.  A random variable can have properties,
and a pair of them can stand in relation to each other---e.g.\ we can ask
if  $X_t$   and  $X_[t+1]$     are independent or not---which is absurd for
variables.)

     Thus, for each instant in time  $t$  we have that the position
of a sphere is governed by a random variable, $X_[t_0]$, which evolves
from the previous random variables   ${X_t}_[-\infty<t<t_0]$      by a law
analogous (seemingly) to a law of motion.  That it is impossible to
consistently adopt this intuitive interpretation is manifest now,
since this is incompatible with that part of the intuitive
interpretation which we exposited previously, \it viz.\rm, that for fixed 
  $\alpha$, the sample path    ${X(t,\alpha)}_[-\infty<t<\infty]$    represents
the motion of the sphere under ``random impacts."  In the one case, we
view  $X(t,\alpha)$    with  variable (unsaturated), in its dependence
on      $t$, ${X_t}$.  That is, we fill in  $t$  first (saturated).  In the other
case, that of viewing the sample path, we fill in $\alpha$   first, and then 
$t$.  These are incompatible.  In particular, the sample path is not a
solution to the stochastic differential equation: e.g.\ every
continuous function (satisfying the initial condition  $X(0)=0$)  is
a sample path.  So the stochastic differential equation imposes \it no
conditions \rm at all on the sample paths.

     In  $V_n$, \it au contraire\rm,  $X(t,v_0)$  is a solution to the
equations of motion no matter whether  $v_0$  is fixed or variable. 
The popular fallacy regarding the Wiener process, exposed above,
has probably arisen through two causes:  Firstly, the false
parallelism between random variables and variables, (most
physicists have never learned the Peano-Frege-Russell analysis of
variables and quantifiers and still think, with Euler and Cauchy,
that there are two types of real numbers:  constant ones, and
variable ones), and secondly, an unjustified carryover of the
interpretation of  $V_n$  to the model  $W$.

     In a sense, of course, the above critique of the naive
interpretation of a stochastic process is essentially the same as
the critique of the naive interpretation of a random variable    
$Y(\alpha)$.  Naively, one thinks of the $Y(\alpha)$       as describing an
outcome, its value, that is random, i.e.\ not determined by anything.  But of
course the linguistic analysis of  $Y(\alpha)$     does not bear this out: 
its value is a function of    $\alpha$, so after all it is deterministic.

     But as long as there was no concrete issue at stake, this
critique was never convincing.  The Wiener process is ``quite
possibly the single most important object in all of 
modern probability theory,"\foo1[Stroock, \it Probability Theory From an Analytic
Point of View\rm, Cambridge, 1993, p.\ xi.]
 but also typical of statistical mechanics. 
The concrete issues discussed above can be regarded, then, as decisive objections
to the naive analysis of a random variable as well as of probabilistic statements.

     Some such view as advanced in this paper is necessary in order
to make sense of von Mises's ensemble or frequency interpretation
of probability statements.  Von Mises himself, in referring a
probability statement (about a single event), for its meaning, to
an (implicit) statement about an ensemble, is using a structurally
similar `unpacking' style of analysis as we have already advanced. 
To make his notion rigorous, it is of course necessary to include
an infinite ensemble and a limiting process (as well as a measure),
as he practically does.  The only step missing from his analysis,
which we supply, is an explicit recognition of the infinite
ensemble (limit)---which is unphysical---as a mathematical
approximation device for making calculations (approximate ones)
about a fixed finite stage in the limiting process.  This amendment
to von Mises is necessary if one wants to insist that meaningful
propositions are about reality, as long as one thinks that reality
is necessarily finite.  That is, talk of potential trials is not
able to fit into the framework of Russellian logic.  But on the
other hand, in practice we see that the actual trials are always a
finite ensemble---even if not always, our theory has to cover the
vast majority of cases where they are finite.  (Naturally, I am not
denying that infinite sets are not used in Russellian logic---but
they are sets of actual objects, never of potential objects.  This
is, in fact, just the same difficulties of random variables all
over again.  If anyone tried to invent a new logic with a new kind
of variable, an $r$-variable which in the interpretation was ``random"
but linguistically was a variable, how could they talk about an
infinite sequence of independent identically distributed trials? 
E.g., a Bernoulli process?  Objects are distinct because they have
different properties.  Random variables are discovered
independent---or not, as the case may be---by comparing their
structure as functions of     $\alpha$.)

     There is another possible objection that can be made.  There is a
widespread misimpression that the success of Quantum Physics has definitively
established stochasticity as a fundamental feature of science, so that it is
perverse to try to explicitly define the notion of probability in terms of
classical logic and classical, deterministic, philosophy of science.  This is
not so.  There has been a recent flourishing of work on the problem of Quantum
Measurement and the old positivistic consensus has completely broken down. 
Advances in Deterministic Chaos and the Gibbs program of statistical mechanics,
even in the quantum context, have opened up the question again, even though
hidden variables is not going to be accepted as the answer.  For this reason,
it is useful to try the experiment of trying to account for the probabilities
that arise in classical thermodynamics etc.\ in a logically careful way.  This
clears the ground for a re-examination of Einstein's question, whether the
probabilities of Quantum Mechanics arise in the same sort of way from
Schroedinger's equation as the probabilities of classical mechanics arise from
the Hamilton-Jacobi equations.  Because Einstein himself and most physicists
have had a philosophically inadequate definition of probability and a sloppy
undergraduate understanding of the logical structure of statistical mechanics. 
A logically careful account of the classical statistical mechanics was first
adumbrated by Darwin and Fowler in the 1920's and never penetrated into the
discourse.  A logically unexceptionable reductionist account of probability is
the accomplishment of this paper.  This experiment will put Einstein's question
in a completely new light.  Since we do not use ignorance, hidden variables, or
coarse-graining at all in our analytical `unpacking' of the structure of
probability assertions, the traditional understanding of the issues posed by
Einstein must be wrong.

Now our analysis of Brownian
motion shows the stochastic aspects can arise as an artifact of the
passage from  $V_n$  to  $W$.   Hence
so, then our claim would be practically justified.  That is,
assertions as to the probability of an individual event have, not
a physical, but a linguistic meaning.  On analysis, they `unpack'
as asserting that such and such an approximation technique yields
such and such an answer, and the approximation is valid---an
assertion which may sometimes fall short of the truth---but only
for exactly the same reasons as other approximations break down,
and not for any reasons especially peculiar to the concept of
probability.  (In particular the failure would not be ``random.")

     Our thesis is related to the analysis of counterfactual
conditionals.  On my view, a counterfactual conditional is
meaningless without a (more or less implicit, in practice)
supplementary dynamical theory of causality, which allows of varied
initial conditions ($v_0$).  To assert the conditional is to assert
that from  $v_0$, the dynamics lead, in theory, to the consequence.

     Our thesis accounts for the probabilistic statements of
ordinary language if such implicit theoretical structures can be
taken to be part of the shared discourse of the speech-act-community in which
the statements are made.  (If the theories are false, we may as well suppose the
statements are false.)  If more than one possible theory could be so regarded, we
have just the typical ambiguities of ordinary language, nothing more.

     I should anticipate one plausible objection before concluding. 
With reference to statistical physics, the validity of the
approximation techniques relies (so far) on the ergodic theorem. 
But, letting  $f$  represent a dynamic variable or observable of the
system, the ergodic theorem only establishes the validity of the
technique for almost all  $f$, i.e.\ all  $f$  except for a set of
measure zero.  Does this instantiate a reliance on a concept of
probability which does not fall under our paradigm, but rather the
popular one?  The popular one would, that is, interpret ``almost
all" as meaning ``probability one," thus one is practically certain
that the approximation is valid for the particular observable  $f$ 
under consideration.

     It need not.  Firstly, the measure space of all  $f$  does not,
generally, have measure one---so, technically, it can not be a
probability space.  Secondly, even if this technical obstacle could
be overcome,  $f$ can not be the argument of a random variable.  The
objection envisions ``validity" as the value of a random variable
whose argument would be  $f$.  But  $f$  is chosen explicitly by the
scientist, so this is clearly not an appropriate model.

     Let  $C$  be the (algebra) of all  $f$.  Let  $A$  be the aforesaid
set of measure zero and  $B = C \smallsetminus A$.  The situation is rather that
at present there is no way to tell whether  $f\in B$         except to
calculate the predictions of the approximation, perform the
experiment, and compare the two.  This is not an unusual situation
in science and so we are not tempted to invoke the popular notion
of probability to explain it.

     Another misunderstanding to be cleared up, one fostered by the
spate of books on chaos theory, is that the probabilistic nature
enters into the models  $V_n$  prior to passage to the limit.  But
this is not true.  The probabilistic assertions require use of
thermodynamic functions such as temperature---now these are only
defined in the limit.  Then, too, physicists, especially those
working on the  $C^*$-algebra approach to statistical fallaciously pass over the
distinction between mixed states and states.  Not only is this distinction
absolutely crucial from a Russell-Wittgenstein point of view, their blurring
it is completely gratuitous when the ergodic theorem is available. 
For in this case, ``phase averages" can be simply a technical device and
need not be given any special physical interpretation.

Now it is time to point out one of the advantages of introducing the passage to
the thermodynamic limit.  We need not in fact assume that the dynamics is
ergodic.  As Khinchine shows, the full strength of the ergodic theorem is not
necessary in the thermodynamic limit.  That is, there will be a wide class of
functions $f$ (only, not necessarily almost all) for which the phase average is
a better and better approximation of the time average, as one proceeds to the
thermodynamic limit \foo1[Khintchine, \it op.\ cit.,\rm New York, 1949,
p.\ 62f.]  We can obviously interpret `probability' in this context, [\it mutatis
mutandis].

     In conclusion, we summarize and exemplify the theses of this
paper.  We assume it is well known that popular, traditional
treatments of probability, such as reichenbach's, von Mises's or the more implicit
(and inchoate) subtext of most physics papers, are linguistically
incoherent and essentially irreconcilable with a Russell-Wittgenstein treatment
of logic.  (Either because of the Rosencrantz and Guildenstern effect, or
because they rely on probability as a primitive notion.)

     But the frequency or ensemble theory itself shows (not
unequivocally), and a careful linguistically-corrected
recapitulation of classical statistical mechanics shows (much more
clearly)---that probability statements arise as features peculiar
to a (concrete) approximation device, without a physical objective
correlative.  Thus ``probability" has no (direct) correlative.

     For example, suppose  $V_N$  is a classical dynamical system, and 
$f$  an observable.  The object is to calculate the time average of 
$f$ over a long but finite time interval.  Instead of this, we
calculate the infinite length time average  $<f>$.  We  hope
that this will be a good approximation to our real object.  The
dynamics of  $V_N$  determines the measure we will use implicitly in
what follows, so we do not, in any sense, interpret the invariant
measure  $d\mu$     on phase space as a probability.  We then
hope that the initial conditions,  $v_0$, of the actual system
studied and  $f$  are such that the ergodic theorem applies, saying 
$<f>$  is equal to the phase average of  $f$,       $\int f\,d\mu$.

     But we cannot calculate  $\int f\,d\mu$, so we use an approximate
technique.  We    embed $V_N$      and  $f$ into a sequence of dynamical
systems  $V_n$ and observables  $f_n$, possessing a thermodynamic limit 
$W$ (and $f_W$).  (Note that  $v_0$  has disappeared.)  This has only a
logical status.  We cannot interpret  $W$  physically, or
intuitively as a limit of the  $V_n$.  Firstly, it is not a dynamic
system at all.  Secondly, although there is a sense in which  $f$ 
can be made variable with  $n$, there is no sense in which  $v_0$  can
be.  Now since  $W$  is stochastic whereas  $V_n$  are deterministic,
this is where probabilities enter into the activity of the
scientist, the model-builder and analyst.

     Then it is a fact that  
  $\int f_n\,d\mu_n$   converges to a limit which can be calculated by
means of  $W$. As
explained in the introduction to Ford-Kac-Mazur, the classical program of
statistical mechanics ever since Gibbs has been to establish this fact as a
theorem, as is happening in a growing number of special cases, such as
those by Ford-Kac-Mazur, Hudson-Parthasarathy, Lewis-Maassen,\foo 1
[Lewis, Maassen, ``Hamiltonian Models of Classical and Quantum Stochastic
Processes,'' [\it Lecture Notes in Mathematics], Berlin, 1984, vol.\ 1055, pp.\
247-276] and many others.  

     The thesis of this paper is that the analytic interpretation, of
unpacking, of these probability statements is preferable to the naive
alternative, to mythologically attribute physical reality to each and every
mathematical manipulation.

A theory of the meaning of probability statements with many similarities to the
present one was put forth by Jan von Plato,\foo 2 [[\it op.\ cit.]] but for some
reason not included in his \instert.  The main difference,
unfortunately for philosophical purposes this is crucial, is that he defines
probability as meaning the time average of a functional.  Whereas we, following
the explicit statement of Wiener\foo 3 [Masani, Wiener, ``Non-linear
Prediction,'' in [\it Probability and Statistics, The Harald Cramer Volume],
ed.\ U.\ Grenander, Stockholm, 1959, p.\ 197: ``As indicated by von Neumann \dots
in measuring a macroscopic quantity $x$ associated with a physical or biological
mechanism\dots each reading of $x$ is actually the average over a time-interval
$T$ [which] may appear short from a macroscopoic viewpoint, but it is large
microscopically speaking.  That the limit $\overline x$, as $T \rightarrow
\infty$, of such an average exists, and in ergodic cases is independent of the
microscopic state, is the content of the continuous-parameter $L_2$-Ergodic
Theorem.  The error involved in practice in not taking the limit is naturally to
be construed as a [\it statistical dispersion] centered about $\overline x$.'' 
Cf.\ also Khintchine, [\it op.\ cit.,] p.\ 44f., ``an observation which gives the
measurement of a physical quantity is performed not instantaneously, but requires
a certain interval of time which, no matter how small it appears to us, would, as
a rule, be very large from the point of view of an observer who watches the
evolution of our physical system. \dots Thus we will have to compare experimental
data  \dots with time averages taken over very large intervals of time.''  And
not the instantaneous value or instantaneous state. Wiener, as quoted in Heims,
[\it op.\ cit.,] p.\ 138f., 
 ``every observation \dots takes some finite time, thereby introducing
uncertainty.''  The finiteness operates in two distinct ways.
  Since it is not instantaneous, it introduces time-averages.
This brings us into the realm of the ergodic theorem, statistical
mechanics, and hence probability.  Since it is not infinite, there
is some error involved in replacing it by infinite-duration time
averages.
  And Benatti, [\it Deterministic Chaos in Infinite Quantum
Sustems,] Berlin, 1993, ``Trieste Notes in Physics,'' p.\ 3, ``Since
characteristic times of measuring processes on macrosystems are greatly longer
than those governing the underlying micro-phenomena, it is reasonable to think
of the results of a measuring procedure as of time-averages evaluated along
phase-trajectories corresponding to given initial conditions.''(underscoring in
the original). And Pauli, [\it Pauli Lectures on Physics, volume 4, Statistical
Mechanics], Cambridge, Mass., 1973, p.\ 28f., ``What is observed macroscopically
are time averages\dots ''] (who was a logician), interpret the time average of a
functional as `measurement.'  Professor von Plato attributes his definition to
Einstein but this must be a mistake.  Einstein's statement to that effect in his
early papers is not to be taken literally as a piece of philosophy, it is
evidently merely a physicist's [\it Ansatz] in order to fudge various irrelevant
difficulties and make room for progress on one difficulty at a time.  Einstein
was not a logician, and his path-breaking work on statistical mechanics preceded
by several decades Darwin's rigourous solution of the difficulties which Einstein
wished to push to one side by means of this [\it Ansatz].

Finally we remark that the structural `jumps' in our unpacking correspond very
well to the naive scientist's idea of the meaning of probability, and this is
evidence in favour of our proposal.  For example, Kolmogoroff, [\it loc.\ cit.,]
analysed the frequency theory of probability as being of rough practical value
nevertheless, in spite of the fact that it could not give a reduction of the
concept of probability to other concepts.  The first feature of this to be noted
is that the frequency ``predicted'' is merely asserted to be close to what will
happen, where close is a vague notion.  In our structural unpacking, this
corresponds to two features: firstly, measurements are in practice always
finite, whereas the ergodic theorem only holds exactly for infinite time
averages.  Modelling measurements by very large time averages is exact, but the
approximation procedure replaces them by infinite time averages, and this
introduces as an approximative vagueness the vagueness which Kolmogoroff thought
must be due to the primitive, undefinability of probability.  Wiener has also
remarked explicitly on the importance of the fact that all our measurements or
observations are of finite time intervals, not infinite ones, and the
unavoidable imprecision this introduces into our predictions.  But our proposal
makes it clear that this vagueness or imprecision is not a logical one, but
merely of the same epistemological category, whatever that is, of traditional
approximate measurements in classical mechanics.  Secondly, the ergodic theorem
holds only for almost all initial conditions.  There is an exceptional set of
measure zero.  Now for finite observations, time-averages over finite time
intervals, the set of initial conditions or trajectories for which the observed
value differs appreciably from the phase average (predicted value) is actually
of positive measure: it shrinks to measure zero only in the limit of infinite
time averages.  So this is the second source of imprecision: the exceptional set
has positive measure.  Our proposal removes this imprecision from the
philosophical realm of probability because of the extra layer of indirection we
have added, when compared to the naive frequency theory.  The statement that the
probability of observing the event is x remains true even when the actual
observation turns out to belong to or be close to the exceptional set and so
leads to a measurement that differs appreciably from the probabilistic
expectation.  Because, according to our proposal, the statement was not directly
about the observation, but was about the modelling procedure leading us to
predict the expectation, and was thus about a whole statistical ensemble of
counter-factual conditionals besides the actual realised observation.  It thus
remains true no matter which conditional is realised.

\vskip -.3in
The second roughness in the naive view has to do with the insufficiently
formalised notion of what is meant physically by saying that the trials are
independent.  This corresponds to the following fold in our unpacking proposal:
we interpret the making of a probability assertion as implicitly making
assertions that there is an underlying dynamics to determine the event and that
this dynamics has various properties (either ergodicity, which, by the way, is
usual in quantum dynamics and easier to establish there than classically, or
some sort of large-number condition which is sufficient for the truth of the
result of the ergodic theorem even when ergodicity is lacking).

Finally, we point out that the third roughness in the usual view, again, rather
acutely formulated by Kolmogoroff, is that even if one sneaky observer somehow
gathered enough initial conditions data about the coin toss to be able to
predict it, we would somehow want to still say that its probability was
objectively or subjectively one-half when the event occurs in the context of a
specification or description that leaves this extra knowledge out of account. 
This is accomplished by the proposal of this paper in the same way as the
explanation of the meaning of independence of trials, above.  That is, the
probability assertion is taken by us as implicitly including a specification of
the entire phase space of the system, including its Liouville measure, indeed,
the phase spaces of each member $V_n$ of an infinite sequence of dynamical
systems.  The statement then being about this sequence, the meaning is left
intact even if we in fact all knew the exact initial condition $v_0$ which
obtained.  The meaning is clear-cut and logical, its connection with
observations and scientific practice is formulatable in precise terms, and yet
the imprecisions we are used to are preserved in the harmless form which they
have in many other places in classical epistemology, quite apart from notions of
probability.

\centerline[\bf Bibliography]

\baselineskip=12pt
\noindent\[1\] Joseph F. Johnson, `Thermodynamic Limits, Non-commutative Probability, and Quantum Entanglement' in, 
Quantum Theory and Symmetries III, Cincinnati 2003, ed. by Argyres et al, Singapore, 2004, pp.133-143.

\noindent\[2\] Jan von Plato, ``Ergodic Theory,'' in Skyrms, Harper, eds., [\it
Causation, Chance, and Credence], [\it Proceedings of the Irvine Conference on
Probability and Causation, 1985] vol.\ 1, Dordrecht, 1988 pp.\ 257-277.  

\end